\documentclass[11pt]{article}
\usepackage[utf8]{inputenc}
\usepackage{listings}
\usepackage{color}
\usepackage{graphicx}
\usepackage{longtable}
\usepackage{float}
\usepackage{amsmath}
\usepackage{amssymb}
\usepackage{authblk}

\title{}
\date{\today}
\begin{document}

\title{Quantum mechanical inspired factorization of the molecule pair propagator in theories of diffusion-influenced reactions}
\author{Thorsten Pr\"ustel} 
\author{Martin Meier-Schellersheim} 
\affil{Laboratory of Systems Biology\\National Institute of Allergy and Infectious Diseases\\National Institutes of Health}
\maketitle
\let\oldthefootnote\thefootnote 
\renewcommand{\thefootnote}{\fnsymbol{footnote}} 
\footnotetext[1]{Email: prustelt@niaid.nih.gov, mms@niaid.nih.gov} 
\let\thefootnote\oldthefootnote
\begin{abstract}
Building on mathematical similarities between quantum mechanics and theories of diffusion-influenced reactions, we discuss how the propagator of a reacting molecule pair can be represented as a product of three factors in the Laplace domain. This representation offers several advantages. First, the full propagator can be calculated without ever having to solve the corresponding partial differential equation or path integral. Second, the representation is quite general and capable of capturing not only the classical Smoluchowski-Collins-Kimball model, but also alternative theories, as is here exemplified by the case of a delta- and step-function potential in one and two dimensions, respectively. Third, the three factors correspond to physical quantities that feature prominently in stochastic spatially-resolved simulation algorithms and hence the interpretation of current and the design of future algorithms may benefit. Finally, the representation may serve as a suitable starting point for numerical approximations that could be employed to enhance the efficiency of stochastic simulations.   
\end{abstract}
\section{Introduction}
\label{sec-1}

Stochastic processes, like Brownian motion, and quantum phenomena can be described by mathematical structures that appear to be quite similar \cite{Roepstorff:1994, Chaichian:2001, Schulman:2005}. A case in point is the transition probability density function (PDF) $p(x, t\vert x_{0}, t_{0})$, a central quantity in the theory of Brownian motion that yields the probability to find a Brownian walker at $x$ at time $t$, provided it started at $x_{0}$ at $t_{0} < t$. The transition PDF can be calculated in terms of a path integral that may formally be written as
\begin{equation}\label{Wiener_Integral}
p(x, t\vert x_{0}, t_{0}) \sim \sum_{x(\tau): x_{0}\rightarrow x} e^{-S_{E}[x(\tau)]},
\end{equation}
where the sum includes all paths $x(\tau)$ that satisfy $x(t_{0}) = x_{0}$ and $x(t) = x$. $S_{E}[x(\tau)]$ denotes a certain functional of $x(\tau)$.
Quite analogously, the Feynmann transition \emph{amplitude} $K(x, t\vert x_{0}, t_{0})$ is also given by a path integral
\begin{equation}\label{Feynman_Integral}
K(x, t\vert x_{0}, t_{0}) \sim \sum_{x(\tau): x_{0}\rightarrow x} e^{\frac{i}{\hslash}S[x(\tau)]},
\end{equation}
where $S[x(\tau)]$ refers to the classical action functional. Although the physical context decribed by Eq.~(\ref{Wiener_Integral}) is quite different from the one represented by Eq.~(\ref{Feynman_Integral}), the path integrals can in general be transformed into each other by means of a Wick rotation \cite{Roepstorff:1994, Chaichian:2001, Schulman:2005} and the functional $S_{E}[x(\tau)]$ in Eq.~(\ref{Wiener_Integral}) is the Euclidean version of the action $S[x(\tau)]$. 

Typically, path integrals come along with partial differential equations (PDE) and integral equations \cite{Wiegel:1975}.

Hence, unsurprisingly, the formal correspondence continues on the level of the associated PDE. As is well-known, both the transition PDF $p(x, t\vert x_{0}, t_{0})$ and the Feynmann amplitude $K(x, t\vert x_{0}, t_{0})$ are the Green's functions (GF) of the diffusion equation 
\begin{eqnarray}\label{Diffusion_Equation}
\frac{\partial p(x, t\vert x_{0})}{\partial t} &=& \bigg[D\frac{\partial^{2}}{\partial x^{2}} + V(x)\bigg]p(x, t\vert x_{0}), \\
p(x, t = 0\vert x_{0}) &=& \delta(x-x_{0}),
\end{eqnarray}
and the Schr\"odinger equation, 
\begin{eqnarray}\label{Schroedinger_Equation}
\frac{\partial K(x, t\vert x_{0})}{\partial t} &=& \bigg[\frac{i\hslash}{2m}\frac{\partial^{2}}{\partial x^{2}} - \frac{i}{\hslash}V(x)\bigg]K(x, t\vert x_{0}), \\
K(x, t = 0\vert x_{0}) &=& \delta(x-x_{0}),
\end{eqnarray}
respectively. 

In this manuscript, we exploit the fundamental relationship between quantum mechanics (QM) and Brownian motion in the context of theories of diffusion-influenced reactions \cite{Goesele:1984, Rice:1985}. In the classical Smoluchowski-Collins-Kimball (SCK) picture \cite{smoluchowski:1917, collins1949diffusion, Goesele:1984, Rice:1985}, the diffusion-influenced reaction of an isolated pair of molecules is decribed by a diffusion equation (Eq.~(\ref{Diffusion_Equation})) with vanishing potential $V(x)$. The actual reaction is instead implemented by imposing suitable boundary conditions (BC) at the so-called encounter radius \cite{smoluchowski:1917, collins1949diffusion, Goodrich:1954, Agmon:1984, kimShin:1999, TPMMS_2012JCP}. However, there is an alternative approach that abandons the BC and postulates a non-vanishing potential \cite{Wilemski:1973, Doi_1:1976, Doi_2:1976, Khokhlova:2012BullKorCS, Isaacson:2014, Prustel_Area:2014, Prustel_Area_General:2014}. This formulation is not only capable of including the classical theory, but it may also serve as a starting point to construct more general theories of diffusion-influenced reactions. At the same time, its mathematical structure is closer to the formalism of QM and hence it may pave the way to borrow techniques from QM for theories of diffusion-influenced reactions.

The manuscript is structured as follows. First, we briefly describe the path decomposition expansion (PDX) that has been applied to the quantum mechanical propagator. Then, we briefly review the theory of diffusion-influenced reactions and discuss the PDX for the classical theory and the Doi (also referred to as volume reactivity) model. We explicitly show that we can recover the GF of an isolated pair from the PDX for the case of the SCK theory (implemented as $\delta$-function potential) in one dimension (1D) and for the volume reactivity model in 2D.
\section{Theory}
\label{sec-2}
\subsection{The Path Decomposition Expansion}
\label{sec-2-1}

The PDX was developed in Refs.~{}\cite{Auerbach:1985} to study quantum tunneling, subsequently it has been used for other applications as well, see Refs.~{}\cite{Halliwell:1993, Schulman:2005}. In Refs.~{}\cite{Yearsley:2008, Yearsley:2009}, this technique was used to calculate the Feynmann propagator for a $\delta$- and step-function potential in an elegant and simple way. We point out that these potentials correspond exactly to the SCK and Doi model in the context of theories of diffusion-influenced reactions, as we will expand on later. 
Here, we consider a quantum mechanical particle in 1D. Then, the propagator satisfies the following convolution relation \cite{Auerbach:1985, Schulman:2005, Yearsley:2008, Yearsley:2009}:   
\begin{equation}\label{qm_propagator_conv}
K(x,t \vert x_{0}) = \frac{i\hslash}{2m}\int^{t}_{0} K(x,t \vert a, \tau) \frac{\partial K^{r}(\xi,\tau  \vert x_{0}, 0)}{\partial \xi}\bigg\vert_{\xi=a}d\tau. 
\end{equation}
This relation means that every path can be split into a restricted part (represented by $K^{r}$) that starts at $x_{0}$ at time $t_{0}=0$, never crosses the boundary at $x=a=0$, but that ends at the boundary at time $\tau$, and an unrestricted part (represented by $K$) that starts at the boundary at time $\tau$ and ends at $x$ at time $t$. 
Similarly, one has 
  \begin{equation}\label{qm_propagator_conv_II}
  K(x,t \vert x_{0}) = -\frac{i\hslash}{2m}\int^{t}_{0} \frac{\partial K^{r}(x,t  \vert \xi, \tau)}{\partial \xi}\bigg\vert_{\xi=a} K(a,\tau \vert x_{0}, 0) d\tau, 
  \end{equation}
which corresponds to a split into into an unrestricted part that starts at $x_{0}$ at time $t_{0}=0$, may cross the boundary $x=a=0$ many times, and finally ends at the boundary after time $\tau$ and a restricted part that starts at the boundary at time $\tau$, never crosses the boundary after that and ends at $x$ at time $t$. 
Combining these two formulas (Eqs.~(\ref{qm_propagator_conv}), (\ref{qm_propagator_conv_II})), one obtains for the Feynmann propagator 
  \begin{eqnarray}\label{qm_propagator_conv_III}
  K(x,t \vert x_{0}) &=& \frac{\hslash^{2}}{4m^{2}}\int^{t}_{0}dT \int^{T}_{0}d\tau \frac{\partial K^{r}(x,t  \vert \xi, T)}{\partial \xi}\bigg\vert_{\xi=a} \nonumber \\
&&\times K(a,T\vert a, \tau) \frac{\partial K^{r}(\xi, \tau  \vert x_{0}, 0)}{\partial \xi}\bigg\vert_{\xi=a}.
  \end{eqnarray}
Refs.~{}\cite{Yearsley:2008, Yearsley:2009} exploit that the three quantities on the rhs of Eq.~(\ref{qm_propagator_conv_III}) can be easier calculated than
the full propagator $K(x,t \vert x_{0})$.
This convolution relation (Eq.~(\ref{qm_propagator_conv_III})) is the central identity that we will use in the following and we will explicitly check that it holds true in theories of diffusion-influenced reactions as well.  
\subsection{Smoluchowski-Collins-Kimball theory}
\label{sec-2-2}

The theory of diffusion-influenced reactions is traditionally described by the diffusion equation (Eq.~(\ref{Diffusion_Equation})) with vanishing $V(x)$, subject to a suitable BC that implements the phyics at the encounter distance. The Collins-Kimball BC \cite{collins1949diffusion}, also referred to as radiation BC, generalizes the Smoluchowski (or absorbing) BC \cite{smoluchowski:1917} and reads
\begin{equation}\label{CK_BC}
D\frac{\partial p_{\text{rad}}(x,t\vert x_{0})}{\partial x}\bigg\vert_{x=a} = \kappa_{a}p_{\text{rad}}(x=a,t\vert x_{0}).
\end{equation}
In the limits $\kappa_{a}\rightarrow\infty$ and $\kappa_{a}\rightarrow 0$, the radiation BC reduces to the absorbing and non-reactive (reflective) BC, respectively. 
The survival probability is defined as
\begin{equation}\label{Survival_Probability}
S_{\text{rad}}(t) = \int^{\infty}_{a}p_{\text{rad}}(x,t\vert x_{0})dx.
\end{equation}
Now we may invoke the diffusion equation (Eq.~(\ref{Diffusion_Equation}), $V(x)=0$) and the definition of the survival probability (Eq.~{}\ref{Survival_Probability}), which tells us that
\begin{equation}\label{timeDerivativeSurvivalProb}
-\frac{\partial S_{\text{rad}}(t\vert x_{0})}{\partial t} = D\frac{p_{\text{rad}}(x,t\vert x_{0})}{\partial x}\bigg\vert_{x=a}. 
\end{equation}
Note that this relation is true independent of the imposed BC at the encounter distance. However, if we impose an absorbing BC, the negative time derivative of the survival probability bears a special meaning. It is the first passage time PDF
\begin{equation}\label{FP_PDF}
f_{\text{FP}}(t\vert x_{0}) := -\frac{\partial S_{\text{abs}}(t\vert x_{0})}{\partial t}.
\end{equation}
Furthermore, in the case of a radiation BC, the quantity $\kappa_{a}p_{\text{rad}}(a, t\vert a)$ is the rebinding time PDF that plays an essential role in understanding spatial stochastic fluctuations \cite{Takahashi:2010p139, Ten_Wolde:2012}. Owing to the Collins-Kimball BC (Eq.~(\ref{CK_BC}) and Eq.~(\ref{timeDerivativeSurvivalProb})
one has  
\begin{equation}\label{Reb_PDF}
f_{\text{reb}}(t) := -\frac{\partial S_{\text{rad}}(t\vert x_{0}=a)}{\partial t} = \kappa_{a}p_{\text{rad}}(a, t\vert a).
\end{equation}
Finally, it will turn out to be convenient to consider the PDF $f_{LR}(\tau\vert x, t)$ for the last reflection time $\tau$ before $t$, given that the molecule is located at $x$ at time $t$, which is closely related to the first passage time PDF via
\begin{equation}\label{LR_PDF}
f_{LR}(\tau\vert x, t) = f_{FP}(t-\tau\vert x). 
\end{equation}

As already mentioned, an alternative and more general approach, that abandons the requirement of a BC at the encounter distance, implements the chemical reaction by adding appropriate sink terms to the equation of motion, i.e. one deals with a diffusion equation featuring a non-vanishing potential (Eq.~(\ref{Diffusion_Equation})).
For instance, the SCK model may be defined by the diffusion equation with $V(x) =- \kappa_{a}\delta(x-a)$,
while for volume reaction theories the potential assumes the form
$V(x) = -\kappa_{r}\Theta(a-x)$.

In the following, we will focus on these two examples and show that the corresponding full propagator can be obtained by a convolution relation that is analog to Eq.~(\ref{qm_propagator_conv_III}).
More precisely, one has
\begin{eqnarray}\label{fundamentalRelation}
p_{V}(x, t\vert x_{0}) &=& \pm D^{2}\int^{t}_{0}dT\int^{T}_{0}d\tau\, \frac{\partial p_{\text{abs}}(x, t-T\vert \xi)}{\partial \xi }\bigg\vert_{\xi=a}\nonumber\\
&&p_{V}(a, T-\tau\vert a)\frac{\partial p_{\text{abs}}(\xi, \tau\vert x_0)}{\partial \xi }\bigg\vert_{\xi=a}.
\end{eqnarray}
Note that if both $x, x_{0}> a$ or $x, x_{0} < a$, one has to add a second term $-p_{\text{abs}}(x,t\vert x_{0})$ on the lhs of Eq.~(\ref{fundamentalRelation}). These two cases also correspond to the positive sign on the rhs, while the cases $x<a, x_{0}> a$ and  $x>a, x_{0} < a$ correspond to the negative sign. Furthermore, Eq.~(\ref{fundamentalRelation}) shows that the restricted propagator $K^{r}$ in the quantum mechanical version (Eq.~(\ref{qm_propagator_conv_III})) corresponds to the GF $p_{\text{abs}}$ with absorbing BC. However, we emphasize that $p_{\text{abs}}$ may describe propagation in the presence of an absorbing boundary \emph{and} a non-vanishing potential.  
\subsection{Partially absorbing trap in one dimension}
\label{sec-2-3}

To check the validity of Eq.~(\ref{fundamentalRelation}), we consider a partially absorbing trap in one dimension at $x=a=0$. Instead of following the classical route via imposing a radiation BC, one may employ the following equation of motion \cite{Datta:1992, Taitelbaum:1992}
\begin{equation}\label{oneDimensionDeltaPotential}
\frac{\partial p_{\delta}(x,t\vert x_{0})}{\partial t} = \bigg[D \frac{\partial^{2}}{\partial x^{2}} - \kappa_{a}\delta(x)\bigg]p_{\delta}(x,t\vert x_{0}).
\end{equation}
The GF solution $p_{\delta}(x,t\vert x_{0})$ has been presented in Ref.~{}\cite{Datta:1992}. Note that this GF is \emph{not} given by exactly the same expression as the solution $p_{\text{rad}}(x,t\vert x_{0})$ of the corresponding radiation BC problem, as detailed in Ref.~{}\cite{Taitelbaum:1992}.
In particular, the GF $p_{\delta}(x,t\vert x_{0})$ is defined for all $-\infty < x, x_{0} < \infty$, in contrast to the SCK solution $p_{\text{rad}}(x,t\vert x_{0})$ that is restricted to the semi-infinite line, $0 < x, x_{0} < \infty$. In the following, we will focus on the case $x<0, x_{0} > 0$, but all other cases can be treated similarly.

We observe that in the $\delta$-function case, the convolution relation (Eq.~(\ref{fundamentalRelation}) may be rewritten in the Laplace domain as product of three prominent waiting time PDF. To see this, we first note that in the present context one has to take into account two different first passage time PDF, corresponding to $x_{0} < 0$ and $x_{0} > 0$. More precisely,
\begin{eqnarray}
f^{<}_{\text{FP}}(t\vert x_{0}) &=& -\frac{\partial S^{<}_{\text{abs}}(t\vert x_{0})}{\partial t}, \quad S^{<}_{\text{abs}}(t\vert x_{0}) := \int^{0}_{-\infty}p^{<}_{\text{abs}}(x,t\vert x_{0})dx, \,\, x_{0} < 0,\qquad \label{S_<}\\
f^{>}_{\text{FP}}(t\vert x_{0}) &=& -\frac{\partial S^{>}_{\text{abs}}(t\vert x_{0})}{\partial t}, \quad S^{>}_{\text{abs}}(t\vert x_{0}) := \int^{\infty}_{0}p^{>}_{\text{abs}}(x,t\vert x_{0})dx, \,\, x_{0} > 0.\qquad \label{S_>}
\end{eqnarray}
Furthermore, one has (see Eqs.~(\ref{timeDerivativeSurvivalProb}), (\ref{FP_PDF}), (\ref{LR_PDF}))
\begin{eqnarray}
f^{<}_{\text{LR}}(T\vert x, t) &=& f^{<}_{\text{FP}}(t-T\vert x) = -D\frac{\partial p^{<}_{\text{abs}}(x, t-T\vert \xi)}{\partial \xi }\bigg\vert_{\xi=a}, \quad x < 0, \label{FP_<}\\
f^{>}_{\text{FP}}(t\vert x_{0}) &=& D\frac{\partial p^{>}_{\text{abs}}(\xi, t\vert x_{0})}{\partial \xi }\bigg\vert_{\xi=a}, \quad x_{0} > 0. \label{FP_>}
\end{eqnarray}
Next, we would also like to point out that, although $p_{\text{rad}}(x, t\vert x_{0})$ is different from $p_{\delta}(x, t\vert x_{0})$ for general $x, x_{0}$, one has
\begin{equation}\label{GF_RAD_DELTA}
p_{\text{rad}}(0, t\vert 0) = 2p_{\delta}(0, t\vert 0),
\end{equation}
and 
\begin{equation}
S_{\delta}(t\vert x_{0}) = S_{\text{rad}}(t\vert x_{0}), \quad x_{0} > 0,
\end{equation}
where
\begin{equation}
S_{\delta}(t\vert x_{0}):= S^{>}_{\delta}(t\vert x_{0}) + S^{<}_{\delta}(t\vert x_{0}) := \int^{\infty}_{a=0}p_{\delta}(x, t\vert x_{0})dx + \int^{a=0}_{-\infty}p_{\delta}(x, t\vert x_{0})dx 
\end{equation}
as discussed in Ref.~{}\cite{Taitelbaum:1992}.
Thus, using Eqs.~(\ref{fundamentalRelation}), (\ref{Reb_PDF}), (\ref{LR_PDF}), (\ref{FP_<}), (\ref{FP_>}) and (\ref{GF_RAD_DELTA}), we can substitute the first passage and rebinding time PDF for the spatial derivative terms and for $\tilde{p}_{\delta}(a, s\vert a)$, respectively, in Eq.~(\ref{fundamentalRelation}) to obtain
\begin{equation}\label{triple_factor_PDF}
\tilde{p}_{\delta}(x,s\vert x_{0}) = \frac{1}{2\kappa_{a}}\tilde{f}^{<}_{LR}(s\vert x) \tilde{f}_{\text{reb}}(s) \tilde{f}^{>}_{FP}(s\vert x_{0}), \quad x < 0, \, x_{0} > 0,
\end{equation}
where we employed the following notation for the Laplace transform of a function $g(t)$
\begin{equation}
\mathcal{L}[g](s):= \tilde{g}(s):=\int^{\infty}_{0}e^{-st}g(t)dt.
\end{equation}

One may also verify Eq.~(\ref{triple_factor_PDF}) explicitly. Transforming the solution given in Ref.~{}\cite{Datta:1992} to the Laplace domain, one arrives at
\begin{equation}\label{GF_delta}
\tilde{p}_{\delta}(x, s\vert x_{0}) = \frac{e^{-v(\vert x\vert+x_{0})}}{2D(v+h)}, \quad x<0, x_{0} > 0,
\end{equation}
where $v:=\sqrt{s/D}$ and $h:=\kappa_{a}/D$.
Then, using the expressions for the GF with absorbing BC in the Laplace domain \cite{carslaw1986conduction}
\begin{eqnarray}
\tilde{p}^{<}_{\text{abs}}(x,s\vert x_{0}) = \frac{1}{2Dv}\bigg[e^{-v\vert x- x_{0}\vert} - e^{-v\vert x + x_{0}\vert} \bigg], \quad x, x_{0} < 0,\\
\tilde{p}^{>}_{\text{abs}}(x,s\vert x_{0}) = \frac{1}{2Dv}\bigg[e^{-v\vert x- x_{0}\vert} - e^{-v(x + x_{0})} \bigg], \quad x, x_{0} > 0,
\end{eqnarray}
we obtain
\begin{eqnarray}
\tilde{f}^{<}_{\text{FP}}(s\vert x_{0}) &=& e^{-\sqrt{s} \frac{\vert x_{0}\vert}{\sqrt{D}}}, \label{FP_neg}\\
\tilde{f}^{>}_{\text{FP}}(s\vert x_{0}) &=& e^{-\sqrt{s} \frac{x_{0}}{\sqrt{D}}}. \label{FP_pos}
\end{eqnarray}
Using Eqs.~(\ref{GF_delta}), (\ref{FP_neg}) and (\ref{FP_pos}), we see that Eq.~(\ref{triple_factor_PDF}) does indeed hold true.
\subsection{Area reactivity model}
\label{sec-2-4}

The second example that we discuss is the area reactivity model in 2D \cite{Khokhlova:2012BullKorCS, Prustel_Area:2014, Prustel_Area_General:2014}, also referred to as Doi model in the case of an irreversible reaction \cite{Doi_1:1976, Doi_2:1976, Isaacson:2014}. The equation of motion reads
\begin{equation}
\frac{\partial p_{\Theta}(r,t\vert r_{0})}{\partial t} = \bigg[D \bigg(\frac{\partial^{2}}{\partial r^{2}} + \frac{1}{r}\frac{\partial}{\partial r}\bigg) - \kappa_{r}\Theta(a-r)\bigg]p_{\Theta}(r,t\vert r_{0}).
\end{equation}
In the following, we will show that 
\begin{eqnarray}\label{convolutionArea}
\tilde{p}_{\Theta}(r,s\vert r_{0})= \pm4\pi^{2}a^{2}D^{2}\,\frac{\partial \tilde{p}_{\text{abs}}(r,s\vert \xi)}{\partial \xi} \bigg\vert_{\xi = a} \, \tilde{p}_{\Theta}(a,s\vert a) \,\frac{\partial \tilde{p}_{\text{abs}}(\xi ,s\vert r_{0})}{\partial \xi}\bigg\vert_{\xi=a} 
\end{eqnarray}
Again, if both $r, r_{0}> a$ or  $r, r_{0} < a$, one has to add a second term $-\tilde{p}_{\text{abs}}(r,s\vert r_{0})$ on the lhs of Eq.~(\ref{convolutionArea}), as in the 1D case discussed before.

Furthermore, we emphasize that, although we consider here explicitly the irreversible reaction, Eq.~(\ref{convolutionArea}) remains valid for the reversible case. In fact, one
only has to substitute $w:= \sqrt{(s+\kappa_{r})/s} \rightarrow \sqrt{(s+\kappa_{r} + \kappa_{d})/(s+\kappa_{d})}$, see Ref.~{}\cite{Prustel_Area_General:2014}. 

To be definite, we consider the case $r_{0} > a$, $r<a$. Then, one has \cite{Prustel_Area:2014, Prustel_Area_General:2014}
\begin{eqnarray}\label{area_GF}
\tilde{p}^{<}_{\Theta}(r,s\vert r_{0}) &&= \frac{I_{0}(wr)K_{0}(vr_{0})}{2\pi a D \mathcal{N}},\\
\mathcal{N} &&= vI_{0}(wa)K_{1}(va) + wI_{1}(wa)K_{0}(va),
\end{eqnarray}
where $I_{0,1}, K_{0,1}$ refer to the modified Bessel function of first and second kind, respectively, and zeroth and first order.\cite[Sec. 9.6]{abramowitz1964handbook}.

Let us now turn to the GF $p_{\text{abs}}(r,t\vert r_{0})$ satisfying absorbing BC. Similarly to the 1-D $\delta$-function case considered before, one has to take into account two different GF $p_{\text{abs}}$, corresponding to the domains $r, r_{0} > a$ and $r, r_{0} < a$.
For $r > a, r_{0} > a$, one has in the Laplace domain
\begin{equation}
\tilde{p}^{>}_{\text{abs}}(r,s\vert r_{0}) = \tilde{p}_{\text{free}}(r,v\vert r_{0}) - \frac{1}{2\pi D}K_{0}(vr)K_{0}(vr_{0})\frac{I_{0}(va)}{K_{0}(va)},
\end{equation}
where
\begin{equation}\label{laplaceFree}
\tilde{p}_{\text{free}}(r, s \vert r_{0}) = \frac{1}{2\pi D}
\biggl\{\begin{array}{lr}
 I_{0}(vr_{0}) K_{0}(vr),&\text{$ r  >  r_{0} $} \\
 I_{0}(vr) K_{0}(vr_{0}), &\text{$r  <  r_{0}$}  
\end{array}
\end{equation}
It follows that
\begin{equation}\label{p_abs_deriv}
2\pi a D \frac{\partial \tilde{p}^{>}_{\text{abs}}(\xi, s\vert r_{0})}{\partial \xi}\bigg\vert_{\xi=a} = \frac{K_{0}(vr_{0})}{K_{0}(va)}.
\end{equation}
Next, for $r < a, r_{0} < a$, one has
\begin{equation}
\tilde{p}^{<}_{\text{abs}}(r,s\vert r_{0}) = \tilde{p}_{\text{free}}(r,w\vert r_{0}) - \frac{1}{2\pi D}I_{0}(wr)I_{0}(wr_{0})\frac{K_{0}(wa)}{I_{0}(wa)}.
\end{equation}
Note that the appearance of $w$ instead of $v$ is due to the fact that within the domain $r < a$ the molecule moves in the presence of a constant potential given by $-\kappa_{r}$. 
Thus, one obtains
\begin{equation}\label{p_abs_deriv_2}
2\pi a D \frac{\partial \tilde{p}^{<}_{\text{abs}}(r, s\vert \xi)}{\partial \xi}\bigg\vert_{\xi=a} = -\frac{I_{0}(wr)}{I_{0}(wa)}.
\end{equation}
Finally, using Eqs.~(\ref{area_GF}), (\ref{p_abs_deriv}) and (\ref{p_abs_deriv_2}), we can verify that indeed Eq.~(\ref{convolutionArea}) does hold true. Similarly, one can show that Eq.~(\ref{convolutionArea}) is fulfilled by the GF corresponding to $r,r_{0} > a$ and $r,r_{0} < a$, as well as by the GF describing the reversible reaction.   

Finally, note that if we used for $r<a$ the expression $-\partial S^{<}_{\text{abs}}(t\vert r)/\partial t$, instead of the flux (Eq.~(\ref{p_abs_deriv_2}) we would \emph{not} recover the full propagator. The reason is that in the case of a $\delta$-function potential the flux is equal to the negative time derivative of the survival probability (up to a sign, cf. Eqs.~(\ref{FP_<}), (\ref{FP_>})). However, in the AR model, due to the $\Theta$-potential, one has instead \cite{Prustel_Area_General:2014} 
\begin{equation}
\frac{\partial S^{<}_{\text{abs}}(t\vert r_{0})}{\partial t} =2\pi a D \frac{\partial p^{<}_{\text{abs}}(r,t\vert r_{0}) }{\partial r}\bigg\vert_{r=a} - \kappa_{r}S^{<}_{\text{abs}}(t\vert r_{0}).
\end{equation}

\section*{Acknowledgments}
This research was supported by the Intramural Research Program of the NIH, National Institute of Allergy and Infectious Diseases. 


\begin{thebibliography}{10}

\bibitem{abramowitz1964handbook}
M.~Abramowitz and I.A. Stegun.
\newblock {\em Handbook of Mathematical Functions with Formulas, Graphs, and
  Mathematical Tables}.
\newblock Dover, New York, 1965.

\bibitem{Isaacson:2014}
I.~Agbanusi and S.A. Isaacson.
\newblock {\em Bull. Math. Biol.}, 76:922, 2014.

\bibitem{Agmon:1984}
N.~Agmon.
\newblock {\em J. Chem. Phys.}, 81:2811, 1984.

\bibitem{Auerbach:1985}
A.~Auerbach and S.~Kivelson.
\newblock {\em Nucl. Phys. B}, 257:799, 1985.

\bibitem{carslaw1986conduction}
H.S. Carslaw and J.C. Jaeger.
\newblock {\em Conduction of Heat in Solids}.
\newblock Clarendon Press, New York, 1986.

\bibitem{Chaichian:2001}
M.~Chaichian and A.~Demichev.
\newblock {\em Path Integrals in Physics}, volume~1.
\newblock IOP Publishing, 2001.

\bibitem{collins1949diffusion}
F.C. Collins and G.E. Kimball.
\newblock {\em J. Colloid Sci.}, 4:425, 1949.

\bibitem{Datta:1992}
P.K. Datta and A.M. Jayannavar.
\newblock {\em Physica A}, 184:135, 1992.

\bibitem{Doi_1:1976}
M.~Doi.
\newblock {\em J. Phys. A: Math. Gen.}, 9:1465, 1976.

\bibitem{Doi_2:1976}
M.~Doi.
\newblock {\em J. Phys. A: Math. Gen.}, 9:1479, 1976.

\bibitem{Goodrich:1954}
F.C. Goodrich.
\newblock {\em J. Chem. Phys.}, 22:588, 1954.

\bibitem{Goesele:1984}
U.M. G\"osele.
\newblock {\em Prog. React. Kinet.}, 13:63, 1984.

\bibitem{Halliwell:1993}
J.J. Halliwell and M.E. Ortiz.
\newblock {\em Phys.Rev.D}, 48:748, 1993.

\bibitem{Khokhlova:2012BullKorCS}
S.S. Khokhlova and N.~Agmon.
\newblock {\em Bull. Korean Chem. Soc.}, 33:1020, 2012.

\bibitem{kimShin:1999}
H.~Kim and K.J. Shin.
\newblock {\em Phys. Rev. Lett.}, 82:1578, 1999.

\bibitem{Ten_Wolde:2012}
A.~Mugler, A.G. Bailey, K.~Takahashi, and P.R. ten Wolde.
\newblock {\em Biophys. J.}, 102:1069, 2012.

\bibitem{TPMMS_2012JCP}
T.~Pr\"ustel and M.~Meier-Schellersheim.
\newblock {\em J. Chem. Phys.}, 137:054104, 2012.

\bibitem{Prustel_Area:2014}
T.~Pr\"{u}stel and M.~Meier-Schellersheim.
\newblock {\em J. Chem. Phys.}, 140:114106, 2014.

\bibitem{Prustel_Area_General:2014}
T.~Pr\"{u}stel and M.~Meier-Schellersheim.
\newblock {\em J. Chem. Phys.}, 141:194115, 2014.

\bibitem{Rice:1985}
S.A. Rice.
\newblock {\em Diffusion Limited Reactions}.
\newblock Elsevier, New York, 1985.

\bibitem{Roepstorff:1994}
G.~Roepstorff.
\newblock {\em Path Integral Approach to Quantum Physics}.
\newblock Springer, Berlin, 1994.

\bibitem{Schulman:2005}
L.S. Schulman.
\newblock {\em Techniques and Applications of Path Integration}.
\newblock Dover, New York, 2005.

\bibitem{Taitelbaum:1992}
H.~Taitelbaum.
\newblock {\em Physica A}, 190:295, 1992.

\bibitem{Takahashi:2010p139}
K.~Takahashi, S.~T{\u a}nase-Nicola, and P.R. ten Wolde.
\newblock {\em PNAS}, 107:2473, 2010.

\bibitem{smoluchowski:1917}
M.~von Smoluchowski.
\newblock {\em Z. Phys. Chem.}, 92:129, 1917.

\bibitem{Wiegel:1975}
F.W. Wiegel.
\newblock {\em Phys. Rep.}, 16:75, 1975.

\bibitem{Wilemski:1973}
G.~Wilemski and M.~Fixman.
\newblock {\em J. Chem. Phys.}, 58:4009, 1973.

\bibitem{Yearsley:2008}
J.M. Yearsley.
\newblock {\em J. Phys. A: Math. Theor.}, 41:285301, 2008.

\bibitem{Yearsley:2009}
J.M. Yearsley.
\newblock {\em Journal of Physics: Conference Series}, 174:012072, 2009.

\end{thebibliography}
\end{document}